\documentclass[]{spie}  
 
\usepackage{amsmath,amsfonts,amssymb}
\usepackage{graphicx}
\usepackage[colorlinks=true, allcolors=blue]{hyperref}

 \title{The Transients Handler System for the Cherenkov Telescope Array Observatory}
 
\author[a]{Kathrin Egberts}
\author[a]{Clemens Hoischen}
\author[a]{Constantin Steppa}
\author[b]{Matthias Füssling}
\author[b]{Dominik Neise}
\author[c]{Emma de Ona Wilhelmi}
\author[\space]{Igor Oya$^\mathrm{b}$ for the CTA Observatory}

\affil[a]{Universit\"at Potsdam, Institut f\"ur Physik und Astronomie, Campus Golm, Haus 28, Karl-Liebknecht-Str. 24/25, 14476 Potsdam-Golm, Germany}
\affil[b]{CTAO gGmbH, Saupfercheckweg 1, 69117  Heidelberg, Germany}
\affil[c]{DESY, Platanenallee 6, 15738 Zeuthen, Germany}

\authorinfo{Further author information: (Send correspondence to K. Egberts)\\K. Egberts: E-mail: kathrin.egberts@uni-potsdam.de}

\begin{document} 
\maketitle
\begin{abstract}
The Cherenkov Telescope Array Observatory (CTAO) will be the largest and most advanced ground-based facility for $\gamma$-ray astronomy. Several dozens of telescopes will be operated at both the Northern and Southern Hemisphere. With the advent of multi-messenger astronomy, many new large science infrastructures will start science operations and target-of-opportunity observations will play an important role in the operation of the CTAO. The Array Control and Data Acquisition (ACADA) system deployed on each CTAO site will feature a dedicated sub-system to manage external and internal scientific alerts: the Transients Handler. It will receive, validate, and process science alerts in order to determine if target-of-opportunity observations can be triggered or need to be updated. 
Various tasks defined by proposal-based configurations are processed by the Transients Handler. These tasks include, among others, the evaluation of observability of targets and their correlation with known sources or objects. This contribution will discuss the concepts and design of the Transients Handler and its integration in the ACADA system.
\end{abstract}


\section{INTRODUCTION}
\label{sec:intro}  
Multi-messenger astronomy makes use of the complementary information accessible via different observational channels. While multi-\textit{wavelength} observations covering the electromagnetic spectrum from radio to very-high-energy $\gamma$-rays with multiple instruments have already been used exhaustively in the past\cite{Boettcher:2019gft}, the utilisation of different \textit{messengers} became feasible only recently\cite{Egberts:20198T} with the improved sensitivity of the current-generation neutrino experiments and the discovery of gravitational waves\cite{LIGOScientific:2016aoc}, leading to the dawn of multi-messenger astronomy with the first multi-messenger observations of a gravitational-wave/short $\gamma$-ray burst correlation\cite{LIGOScientific:2017zic,LIGOScientific:2017ync} and the coincidence of an IceCube-observed neutrino with a flaring blazar\cite{IceCube:2018dnn}. These examples clearly demonstrate the potential of combined observations, which will become even more frequent and impactful with new instrumentation starting operation in the near future\cite{2019ApJ...873..111I, SKA, KAGRA, Adri_n_Mart_nez_2016}. \\ 
The Cherenkov Telescope Array Observatory (CTAO) is the next-generation array of imaging atmospheric Che\-ren\-kov telescopes\cite{CTAConsortium:2013ofs} and is currently about to enter its construction phase. It will operate as a proposal-driven observatory and measure $\gamma$-rays at an energy of tens of GeV to hundreds of TeV. The CTAO features two sites: one in the Southern Hemisphere, close to the Paranal Observatory in Chile, and one in the Northern Hemisphere, at the Roque de Los Muchachos Observatory in La Palma, Spain. The sites will be equipped with dozens of telescopes of three different sizes: the Large, Medium, and Small- Sized Telescopes (LSTs, MSTs, SSTs), resulting in a four-decades coverage in energy with unprecedented instantaneous sensitivity. Its sensitivity for short time scales makes it a powerful instrument for the study of transient phenomena and the variable universe\cite{CTAConsortium:2017dvg}. Correspondingly, transient follow-up observations, such as $\gamma$-ray bursts, gravitational-wave transients, and high-energy neutrino transients are a key science project for CTAO.  
Consisting of pointed telescopes with relatively small fields of view of few degrees, 
the CTAO requires an efficient triggering mechanism that initiates follow-up observations and possibly re-pointing of the telescopes based on additional (often external) information. This is the role of the CTAO Transients Handler.

\section{GENERAL DESIGN}
  \begin{figure} [ht]
   \begin{center}
   \begin{tabular}{c} 
   \includegraphics[height=10cm]{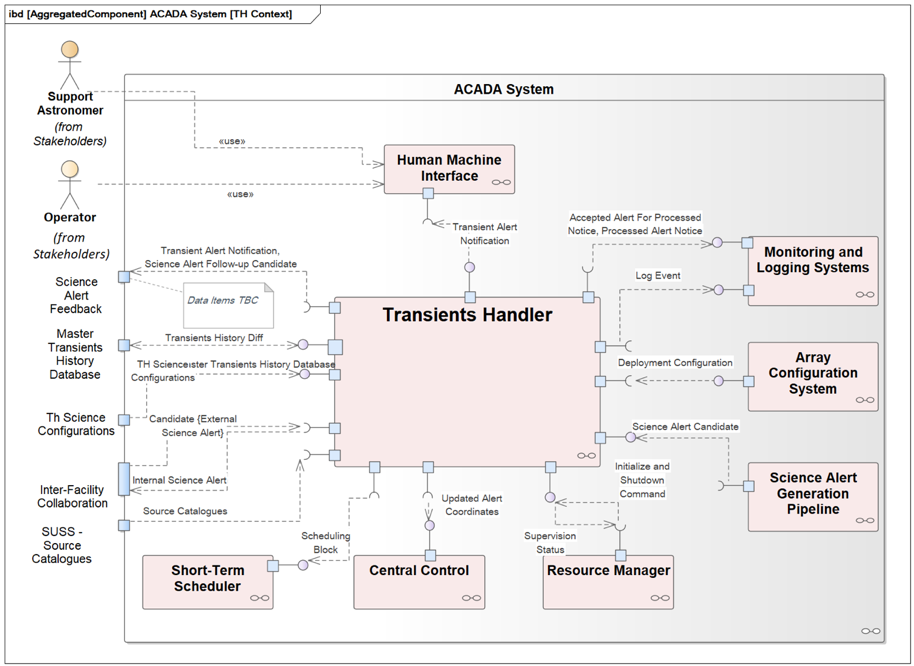}
   \end{tabular}
   \end{center}
   \caption[context diagram of the transients handler] 
   { \label{fig:context} Context diagram of the Transients Handler in UML notation. The main interfaces are described in the text.
}
   \end{figure} 
The Transients Handler is the center piece of the CTAO transients follow-up program. Its task is to receive and manage incoming alerts and process them to arrive at a decision whether or not to observe a transient phenomenon. It is \texttt{python} based and part of the Array Control and Data Acquisition (ACADA) system, which coordinates CTAO observations\cite{oya:hal-03022694}. ACADA is a distributed system and uses the ALMA Common Software (ACS) framework\cite{10.1117/12.461036} as middleware. The role of the Transients Handler in ACADA with its main interfaces to the other ACADA sub-systems as well as its connections beyond ACADA are visualised in Fig.~\ref{fig:context}. 
The Transients Handler receives science alert candidates either from external facilities or internally from the ACADA Science Alert Generation Pipeline (SAG)\cite{Bulgarelli:2021idc}, which performs a real-time analysis of the acquired data, or the Science User Support System (SUSS), which allows for manual triggering of observations. 
Received science alert candidates are processed and evaluated according to predefined criteria (Science Configurations), which are derived from the observation proposals issued to the CTAO. 
Both processing and evaluation of alerts are performed with a maximum availability of the system. 
If a science alert candidate matches the criteria for observations,  
a Scheduling Block is created. It contains the details of the suggested observations and is passed on to the 
Short-Term Scheduler (STS)\cite{2014SPIE.9149E..0HC}, which issues the observation.
Follow-up decisions sent to the STS are one of the following three cases:
\begin{itemize}
\item	Triggering of new follow-up observations,
\item	Updating of previously identified follow-up observations, 
\item	Retracting of previously identified follow-up observations.
\end{itemize}

Since transient science is a rapidly evolving field, the design of the Transients Handler has been chosen to provide a maximum of flexibility. In particular, 
the implementation of the Transients Handler is governed by the design principles of modularity, configurability, and testability. 
\textbf{Modularity} is the most important principle for the design of the Transients Handler. 
A high degree of modularity allows for additions of formats, standards, systems, and technologies in the future. Therefore, a format-agnostic broker system has been chosen that allows to adopt any incoming stream of events in parallel. 
The second most important design choice is a high degree of \textbf{configurability}. Configurability in the handling of transients is needed in order to optimize follow-up decision making and observation planning. 
This is achieved by an extensive Science Configuration data model that allows for detailed configurational changes for individual science cases and the use of science-alert specific processing tasks.
Another important design principle is \textbf{testability}. The processing of science alert candidates is usable independently from the rest of the sub-system 
to perform (i) fine-tuning of Science Configurations, (ii) development of new processing tasks and selection cuts, and (iii) the validation of the alert processing by using reference archival alerts under controlled conditions or self-generated alerts with test Science Configurations that implement all available processing tasks and selection cuts.\\

\label{sec:components}
  \begin{figure} [ht]
   \begin{center}
   \begin{tabular}{c} 
   \includegraphics[height=10cm]{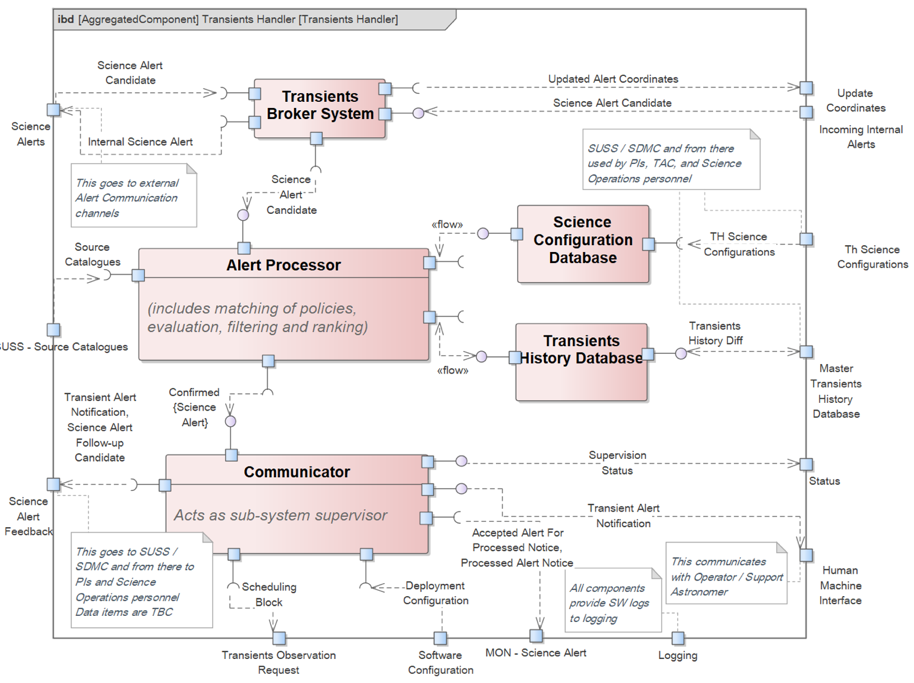}
   \end{tabular}
   \end{center}
   \caption[Component diagram of the Transients Handler. The main interfaces are described in the text.] 
   {Component diagram of the Transients Handler. The main interfaces are described in the text.\label{fig:component} 
}
   \end{figure} 

The Transients Handler consists of three main components, which are visualised in the component diagram of Fig.~\ref{fig:component}: i) the \textbf{Broker System}, providing the entry points for alerts, ii) the \textbf{Alert Processor}, which executes the processing and evaluation of alerts, and iii) the \textbf{Communicator}, which provides information or commands to other sub-systems according to the processing results with the appropriate data format (e.g. Scheduling Blocks for the STS). Base functionality of the Transients Handler is contained in the \textit{Standalone Transients Handler Library}, which has no dependencies on ACADA and ACS, in order to make this part of the Transients Handler usable in CTAO-wide workflows, such as defining and testing Science Configurations, replaying of archival alerts, or even the construction of transients-handling functionality in off-site computing infrastructures.\\ 
The three sub-components are supported by two \texttt{MongoDB} databases, one containing all Science Configurations according to which the received alerts are processed and a second one providing a history of all received and processed science alerts and their aggregated processed information.

\section{The Broker System}

The Broker System proves the interface with internal and external systems in order to receive or send science alerts. The first implemented broker connects with the Gamma-ray burst Coordinates Network (GCN\cite{GCN}), which uses the \texttt{VoEvent2.0}\cite{VOEvent} event format, the current standard of the International Virtual Observatory Alliance (IVOA\cite{IVOA}) for the reporting of transient astrophysical phenomena. To support this format the Transients Handler uses \texttt{comet}\cite{Comet}, which provides all needed functionality for sending and receiving \texttt{VoEvent2.0} alerts.

The modular approach of the Broker System allows to extend the use of the Transients Handler to more formats and sources of alerts. The brokers are implemented as supervised processes, which are managed by the Broker Manager. The Broker Manager interfaces with the Resource Manager\cite{Melkumyan:2019bhp} for central supervision and resource allocation. Changes to the broker's behavior are achieved through adjustments to the configuration of each individual broker. Brokers themselves can pre-select alerts that are allowed to be received - all other alerts will be rejected and not stored.

Three types of brokers are foreseen:
\begin{itemize}
\item	\textbf{External alert brokers}: Brokers that receive alert candidates from networks, brokers, and infrastructures external to the CTAO.
\item	\textbf{Internal alert brokers}: Brokers that receive alert candidates from the SAG or SUSS.
\item	\textbf{Alert broadcasters}: A broker that publishes alerts to other CTAO systems (e.g. SUSS) as well as external systems/networks/brokers. This includes the possibility to pass on information on CTAO observations directly to the outside world. 
\end{itemize}

\section{The Alert Processor}

The Alert Processor is the central component of the Transients Handler. It processes incoming science alert candidates supplied by the Broker System and determines the desired reaction. The number and rates of external alerts depend on the configuration of the Broker System and will change with new experiments coming online. Currently, GCN publishes around 1600 alerts per day. To deal with large numbers of incoming alerts, the alert processing is performed parallelized in multiple threads instead of consecutively.\\
The handling of the science alert candidates is facilitated by the usage of the processing tools of the Standalone Transients Handler Library, which implement the management of alerts, such as I/O from different raw alert formats and the alert validation, as well as the processing according to Science Configurations using processing tasks and post actions in a pipeline approach. Science Configurations specify which processing tasks are executed and what requirements need to be fulfilled to trigger the desired reaction. \\
In a first step, a science alert candidate is interpreted to a common format. The matching with the Science Configurations results in a collection of actual science alerts, where each science alert is a match between the alert candidate and a single Science Configuration (hence multiple science alerts for a single candidate are possible).
The processing of these science alerts is performed in parallel and follows the processing instructions as specified in the Science Configuration.
Also complex processing tasks and post actions can be realised. Processing tasks include all processes that analyse and evaluate the existing and, if necessary, additional external information on an alert. This ranges from the determination of suitable observation time windows to, for example, the calculation of correlation maps based on source catalogues or downloading localisation information for gravitational-wave alerts from the LIGO/Virgo Consortium Data centres, and the determination of an optimized scanning pattern, allowing optimisation for e.g. the best unbiased coverage of the localisation uncertainty or the best coverage of known galaxies within the localisation map\cite{Ashkar_2021}. The optimizations can also feature the sorting of the scanning pattern for e.g. the best zenith angle, or the highest localisation coverage in each position. Based on the results of the processing tasks, the necessary steps are implemented in the post actions, which are required for an actual observation of the target, e.g. the construction of new or updated Scheduling Blocks or notices to invalidate previously constructed Scheduling Blocks.
The activity ends with sending the processed alert to the Communicator and archival in a database.

\section{The Communicator}

The role of the Communicator is to assist with the communication to other ACADA sub-systems. For the core-process of managing science alerts, it provides the functionality to build Scheduling Blocks as well as sending Scheduling Blocks to the STS. Also interfaces to display information on recent and new science alerts in the Human Machine Interface (HMI\cite{HMI}), issuing notifications for the SAG and the Transients-Handler related monitoring are implemented here.

\section{STATUS AND PERSPECTIVES}
\label{sec:sections}
The first version of the Transients Handler incorporates already the science case of follow-up observations of $\gamma$-ray bursts. 
The system is designed to fulfill the requirement of a reaction time of less than 5~s from reception of an alert in the Broker System to the generation of a Scheduling Block (with complex processing algorithms requiring e.g. download of external data like gravitational-wave follow-ups possibly taking longer) and the creation of Scheduling Blocks in less than 1~s.
It is currently in the process of being tested along with the other ACADA components. Towards the end of 2022, ACADA will be integrated with the first LST, LST1\cite{Mazin_2021}, at the La Palma site, with the first tests of having an ACADA-operated telescope running. This also includes the reaction to transient alerts received and processed by the Transients Handler and is a first step towards regular transient follow-up observations of CTAO telescopes initiated by the Transients Handler. 

\section{SUMMARY AND CONCLUSION}
Presented here is the Transients Handler of the CTAO. This ACADA sub-system enables transient science and multi-messenger follow-up observations of the CTAO by serving as an interface between the CTAO observation execution and the (external or internal) triggers for a target of opportunity on short time-scales. The high degree of modularity and configurability of the Transients Handler provides the flexibility that is required to react to changing conditions in the rapidly developing field of multi-messenger astronomy as well as varying observation proposals. A first version of the Transients Handler exists and the integration with the first LST is on-going. With more telescopes to become operational, the Transients Handler ensures that the CTAO can conduct a multifaceted science program dedicated to the transient universe.

\acknowledgments 
    We gratefully acknowledge financial support from the agencies and organizations listed here: http://www.cta-observatory.org/consortium acknowledgments. This work has been funded by the German Ministry for Education and Research (BMBF). 

\bibliography{report} 
\bibliographystyle{spiebib} 

\end{document}